\documentclass[12pt]{article}

 \ifx\pdfoutput\undefined
   \usepackage[dvips]{graphicx}
   \else
   \usepackage[pdftex]{graphicx}
   \fi
   \usepackage{epstopdf}

\newcommand{\be}{\begin{eqnarray}}
\newcommand{\ee}{\end{eqnarray}}

\begin{document}
\DeclareGraphicsExtensions{.pdf,.gif,.jpg}

\bibliographystyle{unsrt}
\footskip 1.0cm

\thispagestyle{empty}
\begin{flushright}
BCCUNY-HEP-07-04
\end{flushright}
\vspace{0.5in}

\begin{center}{\Large \bf {Average $p_t$ as a probe of high energy QCD dynamics}}\\

\vspace{1in}
{\large  Jamal Jalilian-Marian}\\

\vspace{.2in}
{\it Department of Natural Sciences, Baruch College, New York, NY 10010\\ }

\end{center}

\vspace*{25mm}

\begin{abstract}

\noindent Average transverse momentum of hadrons produced in high energy collisions is proposed as a diagnostic probe of high density (saturation) physics. We show that by introducing and varying a transverse momentum cutoff in the definition of the average transverse momentum, one can eliminate the uncertainty in the overall magnitude of the leading order hadron production cross section in high energy collisions, and semi-quantitatively map the different dynamical regions predicted by the saturation physics. We discuss the applications of this method to pion production at RHIC and LHC and make quantitative predictions for the average transverse momentum of produced pions in the kinematics appropriate for the RHIC and LHC experiments.

\end{abstract}
\newpage

\section{Introduction}
Hadron multiplicities and their energy and rapidity dependence were among the first observables measured at RHIC. Contrary to the expectations based on incoherent particle production mechanisms, the hadron multiplicities did not increase with energy as fast as expected. This could be explained using saturation physics and Color Glass Condensate formalism in which coherence plays an essential role. Using a saturation inspired model, Kharzeev, Levin and Nardi \cite{kn} were able to satisfactorily describe the dependence of hadron multiplicities on energy, rapidity and centrality at RHIC. Furthermore, the observed phenomenon of limited fragmentation has a very intuitive explanation in this approach \cite{lim_frag}. Despite the successes of this approach, there are a few issues which need further examination.

A crucial assumption in description of hadron multiplicities in \cite{kn} was the use of the so called parton-hadron duality which was needed in order to turn the produced quarks and gluons into hadrons. Since multiplicities are dominated by low transverse momentum hadrons (for example, about $99\%$ of produced hadrons at RHIC have transverse momenta of less than $1$ GeV or so), one can not make use of the known parton-hadron fragmentation functions which give the probability for a parton to become a hadron. While fragmentation functions are non-perturbative in nature, their dependence on the hard scale can be computed from perturbative QCD to any order in the coupling constant and therefore are a theoretically more robust framework in which to describe hadronization in QCD.

It is also known experimentally that the average transverse momentum of produced hadrons decreases slightly as one goes to more forward rapidities. This would seem to contradict expectations from the saturation physics and Color Glass Condensate (CGC) formalism \cite{glr, nonlin, bk} since in this framework, produced partons have transverse momenta of the order of the saturation scale of the incoming hadrons/nuclei which increases with rapidity. This would imply that the average transverse momentum of produced hadrons should, not only not decrease with increasing rapidity, but rather  increase with increasing rapidity which is not experimentally observed.

Another rather uncomfortable feature of particle production in the CGC formalism is the uncertainty in the overall magnitude of the production cross section \cite{dhj}. This by itself is not necessarily troublesome since all particle production cross sections calculated in the CGC framework have been to the leading order (in $\alpha_s$) accuracy (for work on NLO contributions, see \cite{nlo}). Nevertheless, this uncertainty in the overall magnitude of the production cross section, and the fact that this may be different for a proton target as compared with a gold target somewhat limits the predictive power of the formalism, specially as far as the nuclear modification factor $R_{AB}$ is concerned \cite{rhic, pheno} (for a review of saturation physics and application to RHIC see \cite{ykjjm}). To overcome this, one needs further modeling of the involved cross sections which then further obscures the essential but different dynamics which may be involved in different kinematic regions and for different targets.

It is therefore highly desirable to consider observables which are not sensitive to the above mentioned uncertainties. Here we propose the average transverse momentum, defined with a lower transverse momentum cutoff, of produced hadrons as such an observable. By choosing a (large) lower cutoff in the definition of the average transverse momentum one can minimize the truly non-perturbative aspects of hadronization and utilize the known fragmentation functions of QCD. This alleviates the need for ad-hoc assumptions about the nature of hadronization. Furthermore, this observable is {\it very insensitive} to the uncertainty in the overall magnitude of the  production cross section since the {\it needed $K$ factors are independent of transverse momentum} as shown in \cite{dhj} (another observable which may be insensitive to the $K$ factor is the photon to pion production ratio \cite{phopi}). This enables one to make absolute predictions for particle production in the current and future colliders which is highly needed. 
One can also study the different kinematic regions predicted by CGC in a semi-quantitative way by judiciously choosing the lower transverse momentum cutoff. In principle, one can also address the centrality (or $A$) dependence of the CGC effects in different $p_t$ regions and trace the change in the $A$ dependence of "shadowing" as one varies the transverse momentum. In this work, we focus on the rapidity dependence of the average transverse momentum of produced pions at RHIC and LHC and provide quantitative predictions for different lower cutoffs implemented. 

\section{Average transverse momentum of pions}

In \cite{dhj,hybrid} a hybrid approach to particle production was developed which applies to high energy asymmetric (dilute on dense) collisions, such has deuteron-gold collisions at RHIC and proton-lead collisions at LHC (this approach is also valid for collision of identical objects provided one looks in the kinematic region away from mid rapidity and as long as one does not expect formation of a Quark Gluon Plasma). The essential idea is to treat the incoming dilute
object as a collection of quasi-free partons which evolve according to DGLAP evolution equations while the dense target is treated as a Color Glass Condensate (for the domain of applicability of this approach, we refer the reader to \cite{dhj}) which satisfies the JIMWLK evolution equations (for an alternative formulation which describes the projectile using CGC formalism, see \cite{alt}). 

In this approach, the single inclusive hadron production cross section is given by 
\be
{d\sigma^{p(d) A \rightarrow h(p_t, y_h)\, X} \over d^2 b_t\, d^2 p_t\, d y_h } &=& 
{1 \over (2\pi)^2}
\int_{x_F}^{1} dx_p \, {x_p\over x_F} \Bigg[
f_{q/p} (x_p, Q^2)~ N_F \left(x_g, {x_p\over x_F} p_t , b_t\right)~ D_{h/q} 
\left({x_F\over  x_p}, Q^2\right) +  \nonumber \\
&&
f_{g/p} (x_p, Q^2)~ N_A \left(x_g, {x_p\over x_F} p_t , b_t\right) ~ 
D_{h/g} \left({x_F\over x_p}, Q^2\right)\Bigg]
\label{eq:cs_had}
\ee
where $N_F (N_A)$ is the probability for scattering of a fundamental (adjoint) dipole on the dense target, $f_{q/h}, f_{g/h}$ are the quark (anti-quark) and gluon distribution functions of the incoming dilute hadron and $D_{h/q}$ is the LO pQCD quark-hadron fragmentation function. The factorization scale $Q$ is set equal to the transverse momentum of the produced hadron $p_t$. In case of deuteron-nucleus collisions, we ignore the possible nuclear modifications of the deuteron wavefunction since these are expected to be small in this kinematic \cite{hkn}. Furthermore, the fraction of the incoming target momentum carried by the gluons is denoted $x_g$ and is given by 
$x_g = x_p\, e^{-2 y_h}$. The lower limit in the $x_p$ integration is $x_q^{min}= x_F$ where 
$x_F = {p_t \over \sqrt{s}}\, e^{y_h}$ is the Feynman $x$ of the produced hadron. We use the DHJ parameterization \cite{dhj} of the dipole profiles $N_{F,A}$ which have been successfully used to describe the forward rapidity hadron production data in deuteron-gold collisions at RHIC.
The adjoint dipole scattering cross section is given by (the expression for scattering of a fundamental dipole can be obtained from below by a rescaling of the saturation scale) 
\be  
\label{eq:NA_param}
N_A(r_t,y_h) = 1-\exp\left[ - \frac{1}{4} [r_t^2 Q_s^2(y_h)]^
{\gamma(y_h,r_t)}\right]
\ee
where the anomalous dimension $\gamma$ is given by (for details see \cite{dhj})
\be
\gamma(r_t,Y) &=& \gamma_s + \Delta\gamma(r,Y) \nonumber\\
\Delta\gamma &=& (1-\gamma_s)\,\frac{\log (1/r_t^2\,Q_s^2)}{\lambda\, Y
+\log(1/r_t^2\,Q_s^2) + d\sqrt{Y}}~.  \label{eq:gam_new} 
\ee 
In this work we will only consider minimum bias collisions since the DHJ parameterization was developed and has been tested for these collisions at RHIC. We define the cutoff dependent average transverse momentum of produced pions as 

\be
< p_t > \, \equiv
 {\int_{p_t^{min}} \, d^2 p_t\, p_t \, {d\sigma^{p(d) A \rightarrow \pi^0 (p_t, y_h)\, X} \over d^2 p_t\, d y_h }
\over
 \int_{p_t^{min}} \, d^2 p_t\, {d\sigma^{p(d) A \rightarrow \pi^0 (p_t, y_h)\, X} \over d^2 p_t\, d y_h }  }
 \label{eq:pt_ave}
\ee

\begin{figure}[htbp]
  \begin{center}
   \includegraphics[width=4.5in]{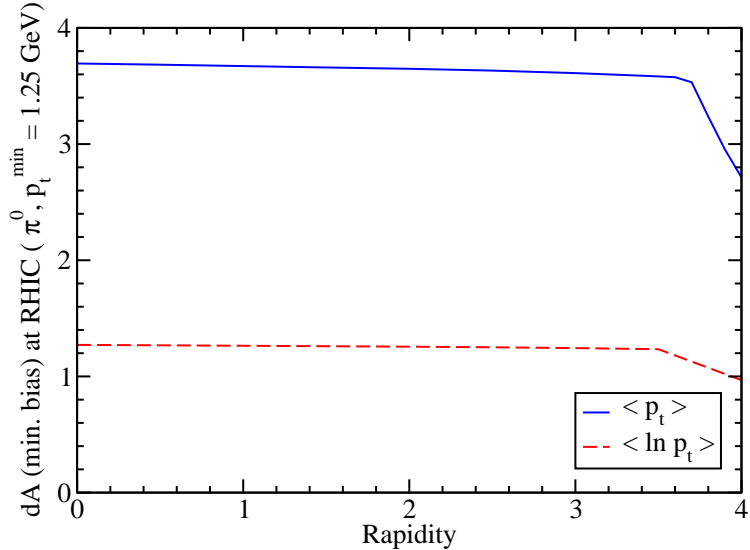}
   \caption{The average transverse momentum of produced neutral pions in min. bias dA collisions at RHIC, with a lower transverse momentum cutoff of $1.25 \, GeV$.}
   \label{fig:r_rhic}
   \end{center}
   \end{figure}

  In Fig. (\ref{fig:r_rhic}) we show the average transverse momentum of produced pions in deuteron-gold collisions at RHIC at different rapidities with a transverse momentum cutoff of $1.25 \, GeV$ which is just about large enough to justify the use of hadron fragmentation functions. Since multiplying the differential cross sections by a power of transverse momentum biases the result toward higher momenta, we also show the average $<ln \, p_t >$ for RHIC. The average transverse momentum of produced pions is slowly decreasing from about $3.7\, GeV$ in mid rapidity to about $3.58\, GeV$ at forward rapidity of $y = 3.5$, a decrease of about $3\%$. Going even more forward results in a faster decrease of the average momentum of produced pions which becomes $2.7 \, GeV$ at rapidity of $y = 4$, a decrease of about $30\%$ from mid rapidity. This fast decrease is due to the shrinking of the available phase space as one goes to very forward rapidity. We have also checked the effect of the DGLAP evolution of the cross section by freezing the factorization scale in (\ref{eq:cs_had}) at $Q_0 = 1.25$ GeV but the effects are minimal and not shown in the figure. A very similar behavior is seen for the average log of transverse momentum. This in principle can be measured at RHIC and will be a very telling signature of whether CGC is the physics responsible for the observed suppression of the pion $R_{dA}$ or some soft beam physics as claimed in \cite{mg}.       

\begin{figure}[hbpt]
   \begin{center}
   \includegraphics[width=4.5in]{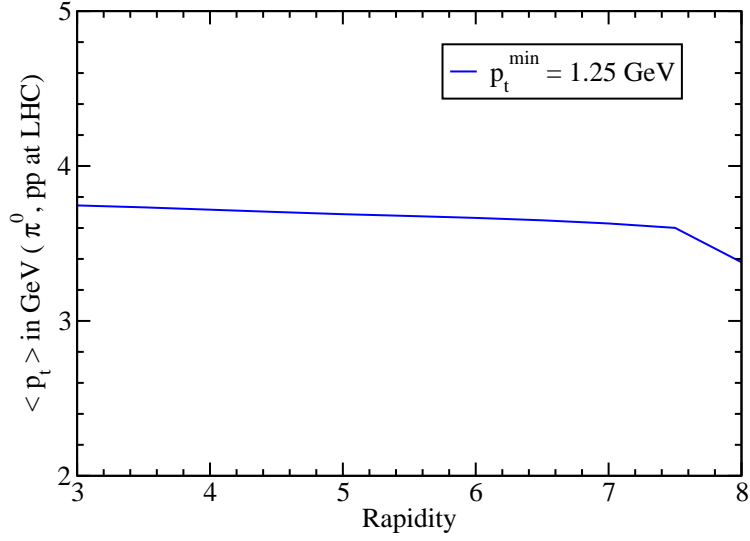}
   \caption{Average transverse momentum of produced neutral pions in proton-proton collisions at LHC ($\sqrt{s} = 14$ TeV).}
   \label{fig:r_pp_lhc}
   \end{center}
   \end{figure}

  In Figure (\ref{fig:r_pp_lhc}) we show the average transverse momentum of neutral pions produced in proton-proton collisions at LHC kinematics ($\sqrt{s} = 14$\, TeV). In order to make our dilute-dense formalism applicable to proton-proton collisions, we consider produced pions only in the forward rapidity, $y \ge 3$, with a transverse momentum cutoff of $1.25 \, GeV$. Going to forward rapidity also helps since the saturation momentum of the target proton may be large enough to justify a CGC approach. Again, a decrease in the average transverse momentum of produced pions is seen as one goes to more forward rapidity. It changes from about $3.7\, GeV$ at $y = 3$ to about $3.6\, GeV$ at $y = 7.5$, a decrease of about $3\%$ while going further in rapidity we see a faster decrease due to the shrinkage of the phase space.  
 
\begin{figure}[hbtp]
   \begin{center}
   \includegraphics[width=4.5in]{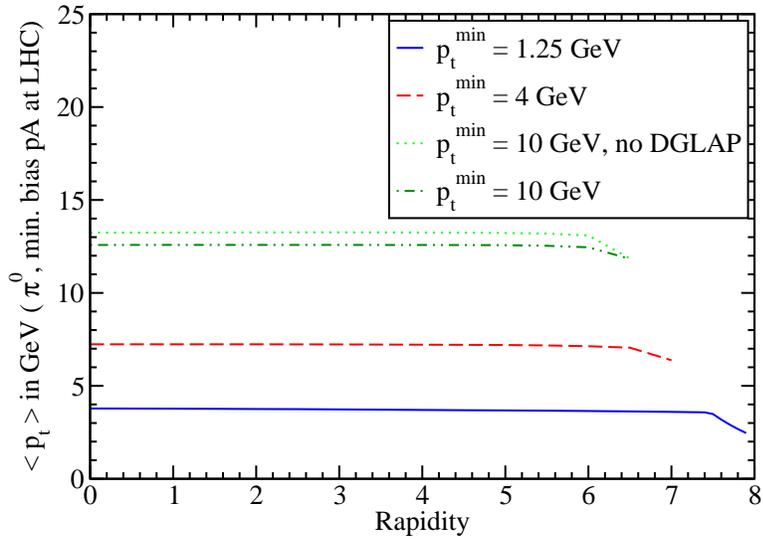}
   \caption{Average transverse momentum of neutral pions produced in min. bias proton-lead collisions at LHC ($\sqrt{s} = 8800$ GeV).}
   \label{fig:r_pA_lhc}
   \end{center}
   \end{figure}

 In Figure (\ref{fig:r_pA_lhc}) we show the average transverse momentum of the produced neutral pions in proton-nucleus collisions at LHC ($\sqrt{s} = 8800$\, GeV) for three different $p_t$ cutoffs. In all three cases, the average transverse momentum decreases slowly with increasing rapidity until very close to the edge of the kinematic phase space, after which it decreases faster. As seen, the approach to the edge of the kinematic limit happens at smaller rapidities as one goes to a higher cutoff in tranverse momentum as expected. In the case where the lower momentum cutoff is $10$ GeV, we also show the effect of freezing the DGLAP evolution of the projectile distribution function as well as the parton-hadron fragmentation function. This corresponds to setting the factorization scale $Q$ in eq. (\ref{eq:cs_had}) to  $Q_0 = 1.25$ GeV. Freezing the factorization scale does not lead to any appreciable change in the average transverse momentum for low cutoffs of $1.25$ GeV and $4$ GeV (not shown) but starts becoming important as one increases the lower cutoff.
By increasing the $p_t$ cutoff the average transverse momentum is pushed to a higher value as seen. This may be useful in mapping out the different kinematic regions predicted by saturation physics. The saturation scale $Q_s$ is the central concept in saturation physics and its absolute magnitude and growth with energy (or $x$) determines the size of the different kinematic regions such as the saturation  and extended scaling regions. The difficulty in determining the saturation scale here is that different values of $x$ contribute at different $p_t$ as one integrates over the transverse momenta of the produced pions. A comparison of the results presented here against experimental measurements of average transverse momentum with different $p_t$ cutoffs would give a better idea about the extent of the kinematic region where CGC approach is valid. The $A$ (or centrality) dependence of our results would also be very useful to investigate, however since we do not have a robust understanding of the $A$ dependence of the dipole profile currently, we leave this for future studies.

To summarize, the average transverse momentum of produced hadrons, with a judicious choice of a lower momentum cutoff, is a very robust and parameter free way of investigating hadron production dynamics  since it can be calculated using the known parton-hadron fragmentation functions. it is also {\it insensitive to the uncertainty in the overall magnitude} of the leading order hadron production cross sections which enables one to make reliable predictions for particle production in high energy heavy ion collisions.

\vspace{0.2in}
\leftline{\bf Acknowledgments} 

\noindent We thank R. Pisarski for discussions related to this topic and F. Gelis for the use of his Fourier transform code.

\vspace{0.2in}
\leftline{\bf References}

\renewenvironment{thebibliography}[1]
        {\begin{list}{[$\,$\arabic{enumi}$\,$]}  % {\arabic{enumi}.}
        {\usecounter{enumi}\setlength{\parsep}{0pt}
         \setlength{\itemsep}{0pt}  \renewcommand{\baselinestretch}{1.2}
         \settowidth
        {\labelwidth}{#1 ~ ~}\sloppy}}{\end{list}}

\end{document}